\documentclass[multphys,vecphys]{svmult}

\usepackage{makeidx}         % allows index generation
\usepackage{graphicx}        % standard LaTeX graphics tool
\usepackage{multicol}        % used for the two-column index
\usepackage[bottom]{footmisc}% places footnotes at page bottom
\makeindex             % used for the subject index
\begin{document}

\title*{Low Luminosity Activity in Hickson Compact Groups}
\author{M. Angeles Martinez\inst{1},
Ascension del Olmo\inst{1}, Jaime Perea\inst{1} \and Roger Coziol\inst{2}}
\institute{Instituto de Astrofisica de Andalucia(IAA),CSIC, Granada (Spain)
\texttt{geli@iaa.es,chony@iaa.es,jaime@iaa.es}
\and Departamento de Astronomia, Universidad de Guanajuato (Mexico)
\texttt{rcoziol@astro.ugto.mx}}

\maketitle

\begin{abstract}
With the aim of studying the influence of environment on the
nuclear activity of galaxies, we have selected a well defined
sample of 65 Compact Groups of galaxies with concordant redshift
in the Hickson Catalogue \cite{mar:H82}.  In this proceeding, we
present the results of the classification of nuclear activity for
42 galaxies, based on newly obtained spectral observations. In
this subsample, 71\% of the galaxies turned out to have emission
lines in their nuclei. 73\% of these emission-line galaxies were
found to have characteristics consistent with low luminosity AGN
(LLAGN), which makes compact groups extremely rich in such
objects.

\end{abstract}

\section{Introduction}

\quad \, Although environment is suspected to play an important
role on the formation and evolution of galaxies, including mass
assembly, star formation, morphological evolution and AGN
activity, the physical details on how such connection is
established are not yet well determined.

To study the connection between environment and nuclear activity
(AGN or HII), systems with high galaxy density and low velocity
dispersion, where the influence of gravitational encounters are
optimized, are in need. In principle, these two conditions can be
found in Hickson Compact Groups of galaxies (HCG). Evidence for
gravitational interaction in these systems take the forms of
morphological and kinematical perturbations \cite{mar:H97}
\cite{mar:Verdes05}, the existence of tidal features such as tails
or shells \cite{mar:Verdes02}, the presence of intergalactic gas
emission observable in X-rays \cite{mar:Ponman96} and the
perturbed distribution of molecular and  atomic gas
\cite{mar:Yun97}.

For our study of nuclear activity we have selected all the groups
with concordant redshift in the Hickson catalogue \cite{mar:H82}
with $\mu_{G}\leq24.4$ (group compactness) and $z\leq0.045$
(redshift completeness), which resulted in a statistically
complete sample of 65 groups(283 galaxies).

% use \sectionmark{}
% to alter or adjust the section heading in the running head

\section{New observations and data reduction}

\quad \, New optical spectroscopic data for 42 galaxies were
obtained with the CAFOS spectrograph, during a seven night run
(November 2004) on the 2.2m telescope of the Calar Alto
Observatory (Almeria, Spain). The detector used was a 2048x2048
SITE CCD, with a plate scale of 0.53arcsec/pixel. Specifications
for the two grisms used to obtain the required wavelength range
can be found in Table 1.

\begin{table}
\centering
\caption{Instrumental parameters}
\begin{tabular}{cccc}
\hline\noalign{\smallskip}
Grism\quad \, & Dispersion\quad \, & Resolution\quad \, &Spectral Range\quad \,\\
\noalign{\smallskip}\hline
B100      & 2\AA/px    &3.74\AA/arcsec &    3200\AA-5800\AA  \\
G100     &  2.12\AA/px  &3.97\AA/arcsec  &   4900\AA-7800\AA  \\
\noalign{\smallskip}
\end{tabular}
\end{table}

The spectra were reduced using standard procedures in IRAF and
SIPL packages.  After overscan and bias subtraction, flat fielding
correction was performed using internal flats (taken with a quarz
lamp). A correction for the different illumination along the slit
 was done using sky flats.
Finally, cosmic-rays were removed by a median combination of
different images (at least three). During each night we also
obtained arc (HgCdAr, He and Rb) expositions for wavelength
calibration and observed three blue and three or four red
spectrophotometric standard stars for flux calibration. Using our
observation of standard stars, an extinction curve was determined
for each night, which resulted in photometric errors less than 5\%
on the calibrated fluxes. The sky contribution was calculated by
selecting two bands on both sides of the target galaxy. After sky
subtraction we made the alignment of the spectra along the spatial
axis.  In figure 1 and 2 we show some examples of spectra obtained
after addition of the three central spatial sections of each
galaxy and after joining the spectra obtained with the two grims.

\begin{figure}
\centering
\includegraphics[height=5.8cm,angle=-90]{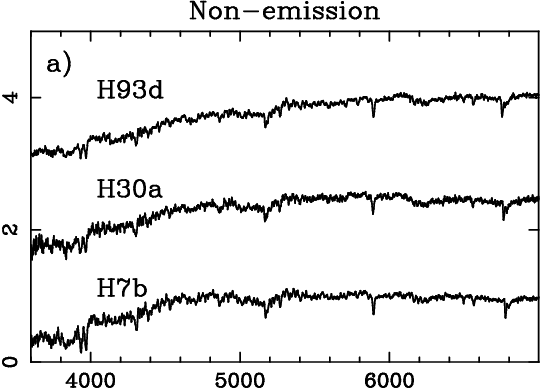}
\includegraphics[height=5.8cm,angle=-90]{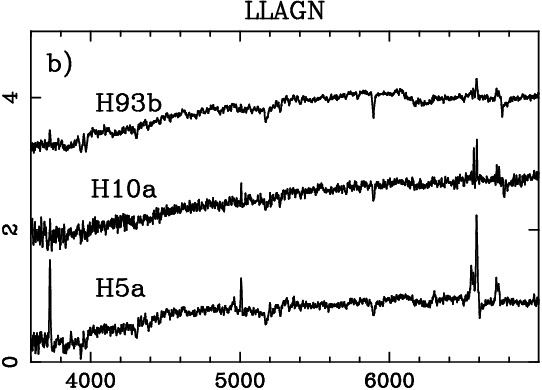}
\caption{Some examples of our new spectra. The flux scale is
  normalized to 1 and the spectra are shifted for ease of presentation.}
\end{figure}

\begin{figure}
\centering
\includegraphics[height=5.8cm,angle=-90]{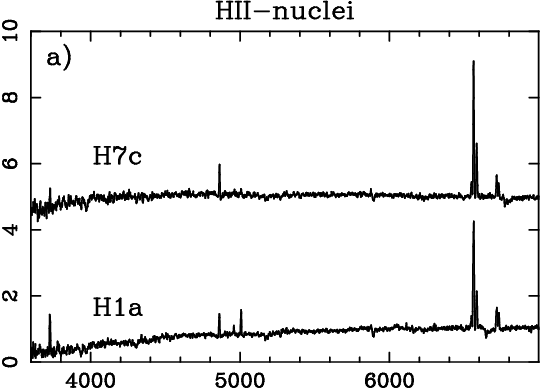}
\includegraphics[height=5.8cm,angle=-90]{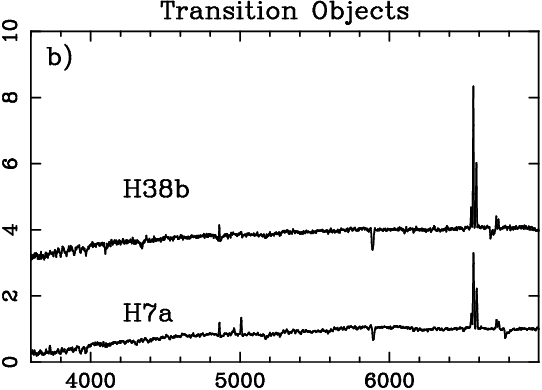}
\caption{Some examples of emission line galaxies.}
\end{figure}

\section{Results and analysis}

\quad \,The classification of activity type of the emission line
galaxies was done using standard diagnostic diagrams
(\cite{mar:BPT81}, \cite{mar:V87}). For our classification, we use
the line ratios log([OIII]/$H_\beta$) and log([NII]/$H_\alpha$),
which are not affected by reddening and are largely insensible to
errors in the photometry. In figure 3 we plot the diagnostic
diagram for all the analyzed objects (the 42 spectra obtained in
Calar Alto together with 68 galaxies previously classified by
Coziol et al. \cite{mar:C98}\cite{mar:C04}). In this figure we
uses the empirical separation between starburst galaxies and AGNs
as determined by Veilleux and Osterbrock \cite{mar:V87}, and the
separation defined by Coziol \cite{mar:C96} to distinguish between
Seyfert and LINERs. Note that in cases where only one line ratio
is available, we tentatively classify the object as AGN based on a
value of [NII]/$H_\alpha$ $>$ 0.6 (\cite{mar:Ho93}).

To summarize our classification, we have found four different
types of spectra (as shown in figure 1 and 2):
\begin{itemize}

\item Non-emission galaxies, where only stellar absorptions features such
   as the CaII lines, G band, 4000\AA \, break or Mg band are observed (Fig
   1a).\\

\item Galaxies hosting  nuclear AGN activity, mainly in the form of Low
   Luminosity AGNs. In figure 1b, emission lines in HCG 93b are obviously diluted
   by intermediate age stellar population which leaves visible only the [NII] lines.
   In HCG 10a, on the other hand, the emission lines are more intense and the
   stellar contribution probably lower, but the intensity of [NII] lines are also quite strong, comparable
   to the intensity of $H_\alpha$, which also suggest an AGN nature. At it was previously shown
   (\cite{mar:C98}\cite{mar:C04}) such spectral characteristics are
   typical of LLAGN, and usually  remain even
   after subtracting a stellar contribution from their spectra.\\

\item Galaxies with star formation in their nuclei, where, as we can
  see in Figure 2a, the strong emission lines are detected over a
  featureless continuum.\\

\item Transition objects between HII/LINER (Fig 2b).\\

\end{itemize}

\begin{figure}
\centering
\includegraphics[height=8.5cm,angle=-90]{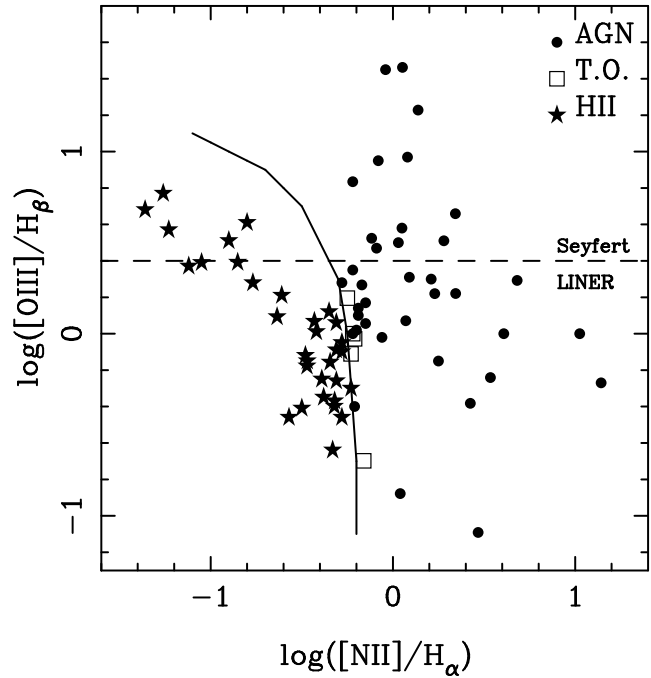}
\caption{Diagnostic diagram for  the
  classified objects. The solid lines indicates the
empirical separation between starburst galaxies and AGNs, as
determined by Veilleux \& Osterbrock \cite{mar:V87}. The dashed line
establishes a distinction between Seyfert2 galaxies and LINERs as
determined by Coziol\cite{mar:C96}.}

\end{figure}

In diagnostic diagrams, transition objects (TO; open squares in
our diagnostic diagram) are located at the limit between LINER and
HII-nuclear region, and are consequently thought to share features
proper to both types. For instance, they show $H_\alpha$ lines
which are more intense than the [NII] lines, although not as
intense as in HII-nuclear galaxies. They also show a continuum
which is not strong enough to mask emission lines. It has been
suggested by Ho et al. \cite{mar:Ho93} that these galaxies could
be the host of an AGN (LINER or Seyfert 2) with a circumnuclear
star formation region. Our present classification of these objects
should be considered temporary, until a proper stellar population
subtraction is performed.

It is interesting to note that in the whole subsample classified
here, we find only one galaxy, HCG 5a (in Figure 1b), with a wide
component in $H_\alpha$ and $H_\beta$ characteristic of a Seyfert
1.5.

\section{Conclusions}

\quad \, Of the 42 galaxies studied in this work, only 8 galaxies
do not show any apparent emission. In 30 galaxies emission-lines
are clearly observed, while a weak presence of [NII] or $H_\alpha$
is detected in four more. The percentage of emission line galaxies
in our present sample amount therefore to at least 71\%. Within
the 30 emission galaxies in our sample, 8 are classified as
HII/star-forming nucleus, 17 (57\%) show characteristics
consistent with low luminosity AGNs with a LINER type, while the
rest (5 galaxies) could represent transition like objects or TO.
If TO really turn out to possess an AGN in their nucleus, that
would increase the number of AGNs in our sample to 73\%.

If we add to our sample the 68 galaxies previously studied by
Coziol et al. (\cite{mar:C98}\cite{mar:C04}) the fraction of
emission-line galaxies would amount to 67\% and 60\% of these
would be classified as AGN (mostly LLAGN). From these results we
conclude therefore that compact groups of galaxies are
particularly rich in AGNs.

%\begin{table}
%\centering
%\begin{tabular}{lll}
%\hline\noalign{\smallskip}
%\subsection{Subsection Heading}
%\label{sec:2}

%\begin{figure}
%\centering

%\includegraphics[height=4cm]{figure.eps}

%\picplace{5cm}{2cm} % Give the correct figure height and width in cm
%\end{figure}
%
% BibTeX users please use
%\bibliographystyle{}
%\bibliography{}{

% Non-BibTeX users please follow the syntax
% the syntax of "referenc.tex" for your own citations
%%%%%%%%%%%%%%%%%%%%%%%% referenc.tex %%%%%%%%%%%%%%%%%%%%%%%%%%%%%%
% sample references
% "physics"
%
% Use this file as a template for your own input.
%
%%%%%%%%%%%%%%%%%%%%%%%% Springer-Verlag %%%%%%%%%%%%%%%%%%%%%%%%%%

%
% BibTeX users please use
%\bibliographystyle{}
%\bibliography{}


%%%%%%%%%%%%%%%%%%%%%%%%%%%%%%%%%%%%%%%%%%%%%%%%%%%%%%%%%%%%%%%%%%%%%%  }

%%%%%%%%%%%%%%%%%%%%%%%%%%%%%%%%%%%%%%%%%%%%%%%%%%%%%%%%%%%%%%%%%%%%%%

\printindex

\end{document}